\documentclass[12pt]{article}

\usepackage{amsmath,amssymb,amsthm}
\usepackage[margin=2.5cm]{geometry}
\usepackage{hyperref}
\usepackage{color}
\usepackage{graphicx}
\usepackage{pgf}

\newcounter{Theorems}
\newcounter{Definitions}
\newcounter{Conjectures}
\begin{document}
\begin{titlepage}
\begin{flushright}

\end{flushright}

\begin{center}
{\Large\bf $ $ \\ $ $ \\
B-RNS-GSS formalism and $L_{\infty}$-actions
}\\
\bigskip\bigskip\bigskip
{\large Andrei Mikhailov}
\\
\bigskip\bigskip
{\it Instituto de Fisica Teorica, Universidade Estadual Paulista\\
R. Dr. Bento Teobaldo Ferraz 271, 
Bloco II -- Barra Funda\\
CEP:01140-070 -- Sao Paulo, Brasil\\
}

\vskip 1cm
\end{center}

\begin{abstract}
Pure spinor formalism and RNS formalism are related by a chain of equivalences constructed
by introducing and integrating-out BRST quartets. This is known as B-RNS-GSS formalism.
The first step is to add BRST quartets to the RNS model and do a similarity transformation
on the space of fields. We show that this step  can be understood as a strictification
procedure which lifts a strong homotopy action of the supersymmetry algebra
to a quasiisomorphic strict action.
This observation allows us to clarify the details of the B-RNS-GSS derivation.
We obtain a closed formula for the similarity transformation as a path-ordered
exponential, and derive the supersymmetric currents.
\end{abstract}

\vfill
{\renewcommand{\arraystretch}{0.8}%
}

\end{titlepage}

\tableofcontents

\section{Introduction}\label{Introduction}

Both RNS and pure spinor string perturbation theories have difficulties at higher loops.
However, the way those difficulties  manifest themselves in those two formalisms is very different.
Thus we may hope that the relation between the two formalisms may help to resolve both.
Therefore, it is desireable to understand the relation, established in \cite{Berkovits:2021xwh},
from various points of view. In this paper we will look at it from the point of view of supersymmetry.
An important ingredient of \cite{Berkovits:2021xwh} is a similarity transformation
relating RNS model with added BRST quartets to the B-RNS-GSS model.
Here we will show that this similarity transformation can be understood in the general context of
{\bf strictification} of $L_{\infty}$ structures
(see Part II of \cite{KrizMay1995} for the general description).

Strictification means constructing a larger space adding auxiliary fields
(``BRST quartets'') so that the resulting space has a strict action of the symmetry algebra.
Integrating out these auxiliary fields results in the original $L_{\infty}$-action.
In particular, there is a functorial correspondence between $L_{\infty}$-actions of Lie superalgebras
and the strict (``usual'') actions, which has been developed in the case of
supersymmetry algebra in \cite{Elliott:2022brv}.

We show that
there is an $L_{\infty}$-action of the Lie superalgebra of supersymmetries on the
large Hilbert space of RNS superstring, and its  strictification
is the B-RNS-GSS construction of \cite{Berkovits:2021xwh}.
The strictification is not unique.
The procedure of \cite{Berkovits:2021xwh} is different from \cite{Elliott:2022brv}, although similar.
The difference is that the construction of \cite{Elliott:2022brv} has finite-dimensional space
of auxiliary variables, while in \cite{Berkovits:2021xwh} the auxiliary variables are holomorphic
fields of a curved beta-gamma system. This is necessary because the resulting model
is required to be a well-defined two-dimensional CFT.

We explain
in Section \ref{GeneralAbsoluteSimilarityTransformation},
that the strictification procedures of the type considered in \cite{Elliott:2022brv}
can be understood as some similarity transformation.
We then develop the field theory analogue of this  similarity transformation, when
the auxiliary variables are holomorphic operators of the chiral de Rham complex. We show that
the resulting procedure is a refinement of the similarity transformation considered in \cite{Berkovits:2021xwh}.
We write an explicit expression for this similarity transformation as a path-ordered exponential,
and explain in which sense the RNS supersymmetry generators are mapped
to the B-RNS-GSS supersymmetry generators.

In the RNS model, the symmetries act on the space of quantum operators of the theory,
but there is no manifest action on the fields in the path integral.
In this sense, the symmetries only manifest after the quantization.
In the ``strictified'' B-RNS-GSS model the supersymmetries actually just act on the fields under
the path integral. We can almost say that they are just geometrical symmetries of the target space
\cite{Bernardes:2023nit}.

\section{B-RNS-GSS construction as strictification}\label{BRNSGSS}

\subsection{$L_{\infty}$-actions and strictification}\label{sec:GeneralStrictification}

Physical quandities are invariant under global symmetries.
At the same time, the Lagrangian description usually has redundancy.
To obtain physical quantities, one has to first pass to the cohomology of the BRST operator.
Before we pass to cohomology, the symmetries still act, but they typically only close up to
BRST-exact transformations. At this level, it turns out that the notion of Lie algebra of
symmetries is not the most economical description. Instead of a Lie algebra
\begin{equation}[t_a,t_b] = f_{ab}^c t_c\end{equation}
it is better to consider a nilpotent fermionic vector field \cite{Alexandrov:1995kv}:
\begin{equation}\label{DifferentialOperator}{1\over 2} f_{ab}^c C^aC^b {\partial\over\partial C^c}\end{equation}
where $C^a$ are coordinates on the supermanifold $\Pi \mathfrak{g}$ --- the superspace of Lie algebra
with opposite statistics \cite{BernsteinLecture1}.
If $\mathfrak{g}$ is an ordinary Lie algebra (as opposed to superalgebra), then all $C^a$ are
fermions. They are very similar to the Faddeev-Popov ghosts in the quantization of gauge theories,
except here they are not fields but constants. We will call them ``background ghosts''.
Eq. (\ref{DifferentialOperator}) defines a differential operator acting on smooth functions
on $\Pi\mathfrak{g}$. The space of smooth functions is traditionally denoted:
\begin{equation}C^{\infty}(\Pi \mathfrak{g})\end{equation}
In this paper we will often consider formal Taylor series of $C^a$,
which often truncate to polynomials. To make the text more readable, we will still use the notation
$C^{\infty}(\Pi \mathfrak{g})$. When a function is only a formal Taylor series, we say this in words.

In this language, a representation $V$ of $\mathfrak{g}$ corresponds to a
nilpotent fermionic operator:
\begin{equation}\label{IntroStrictAction}{1\over 2} f_{ab}^c C^aC^b {\partial\over\partial C^c}
+ C^a \rho_a\end{equation}
where $\rho_a$ are linear operators acting in $V$.
It is often more straightforward to consider differential operators
of the form Eq.  (\ref{IntroStrictAction}), than to consider Lie algebras and their representations.
Especially in the case of Lie superalgebras, when the traditional definition requires
separating coordinates into even and odd and considering separately commutator and anticommutator.
Identifying  Lie superalgebra as a quadratic odd nilpotent vector field results in a more streamlined definition.

From this point of view,
the notion of $L_{\infty}$-action is a direct generalization of the notion of representation.
Instead of Eq. (\ref{IntroStrictAction}), we postulate \cite{Lada:1994mn}:
\begin{align}
d_{L_{\infty}} \;\in\; & \mbox{End}\left(C^{\infty}(\Pi\mathfrak{g})\otimes V\right)\nonumber  \\
d_{L_{\infty}} \;=\; & {1\over 2} f_{ab}^c C^aC^b {\partial\over\partial C^c}  + \underset{0}{q} + \underset{1}q{}_a C^a +
\underset{1}q{}_{ab} C^a C^b + \ldots\label{IntroLinfty}\end{align}
In physics, $V$ is the space of states and $\underset{0}q$ is the BRST operator.
The structure defined by Eq. (\ref{IntroStrictAction}) is
called ``strict action'', to distinguish it from more general Eq. (\ref{IntroLinfty}).

Given a non-strict $L_{\infty}$-action,  we can always lift it to a strict action, by adding some ``BRST quartets''.
This  replaces $V$ with a larger space $\widetilde{V}$, with the differential of the strict form,
like in Eq. (\ref{IntroStrictAction}):
\begin{align}
d_{\rm strict}\;\in\; & \mbox{End}\left(C^{\infty}(\Pi\mathfrak{g})\otimes \widetilde{V}\right)\nonumber  \\
d_{\rm strict} \;=\; & {1\over 2} f_{ab}^c C^aC^b {\partial\over\partial C^c} +
\underset{0}{\widetilde{q}} + C^a\underset{1}{\widetilde{q}}{}_a\label{IntroDStrict}\end{align}
where $\underset{0}{\widetilde{q}}$ and $\underset{1}{\widetilde{q}}{}_a$
are linear operators in $\widetilde{V}$, such that it is a sum of two anticommuting differentials:
\begin{align}
d_{\rm strict}\;=\; & d_0 + d_1\label{IntroBicomplex}\end{align}
with the following properties:

\begin{itemize}
\item The ``UV'' differential $d_1$ commutes with the multiplication by functions of $C^a$,
{\it i.e.}:
\begin{equation}d_1\in \mbox{End}_{C^{\infty}(\Pi \mathfrak{g})}\left(C^{\infty}(\Pi\mathfrak{g})\otimes \widetilde{V}\right)\end{equation}
\item Consider the cohomology $H(d_1)$ of $d_1$, which is the subspace of
$C^{\infty}(\Pi\mathfrak{g})\otimes\widetilde{V}$ annihilated
by $d_1$ modulo the image of $d_1$. There is a homotopy retract (see \cite{Vallette2012} for a review):
\begin{equation}\label{IntroHomotopy retract}\left(\,C^{\infty}(\Pi\mathfrak{g})\otimes\widetilde{V}\,,\;d_0 + d_1\,\right)
\underset{i}{\overset{p}{\rightleftharpoons}}
\left(\,C^{\infty}(\Pi\mathfrak{g})\otimes H(d_1),d_0\right)\end{equation}
where all maps commute with the multiplication by functions of $C^a$.
\item The induced action of $d_0$ on $C^{\infty}(\Pi\mathfrak{g})\otimes H(d_1)$ is equivalent to Eq. (\ref{IntroLinfty}),
{\it i.e.} there is an isomorphism of $C^{\infty}(\Pi \mathfrak{g})$-modules:
\begin{equation}\label{IntroIsomorphism}I\;\in\;\mbox{Hom}_{C^{\infty}(\Pi\mathfrak{g})}
\left(\,
      C^{\infty}(\Pi\mathfrak{g})\otimes V\,,\;
      C^{\infty}(\Pi\mathfrak{g})\otimes H(d_1)
      \,\right)\end{equation}
intertwining $d_{L_{\infty}}$ with the induced action of $d_0$.

\end{itemize}

This implies that the $L_{\infty}$-action $(V,d_{L_{\infty}})$ is quasiisomorphic
to the strict action $(\widetilde{V},d_{\rm strict})$.

\subsection{$L_{\infty}$-action of supersymmetry in RNS model}\label{sec:IntroRNS}

The RNS  formalism for the superstring is based on
worldsheet supersymmetry. The worldsheet fields consist of bosonic coordinates
$X^m$ ($m = 0,\ldots,9$), their fermionic superpartners $\psi^m$, and the
ghost system $(b,c)$ for reparametrization invariance together with the
superghosts $(\beta,\gamma)$ for worldsheet local supersymmetry.

It is convenient to fermionize the superghost system:
\begin{equation}\gamma = e^{\phi}\eta\,,\qquad \beta = e^{-\phi}\partial\xi\end{equation}
where $\eta$, $\xi$ are fermionic fields and $\phi$ is a chiral boson:
\begin{equation}\langle \phi(z) \phi(0)\rangle = -\mbox{ln}(z)
\,,\quad
e^{\phi(z)} e^{\phi(0)} = {e^{2\phi}\over z} + \mbox{regular}\end{equation}
Spacetime supersymmetry generators can be constructed \cite{Friedan:1985ge} as:
\begin{equation}\underset{1}q{}_{\alpha} = \oint dz\; e^{-\phi/2}\Sigma_{\alpha}\end{equation}
where $\Sigma_{\alpha}$ is the spin field of conformal weight $5/8$ constructed
from $\psi^m$. Space-time translations are generated by:
\begin{equation}\underset{1}q{}_m = \oint dz\; \partial X(z)\end{equation}
The {\it large Hilbert space} includes the zero mode of $\xi$, and one defines
the extended BRST operator:
\begin{equation}Q'_{\rm RNS} = Q_{\rm RNS} + \oint dz\,\eta(z)\end{equation}
Notice that $\xi e^{-\phi}$ can be considered ``$\gamma^{-1}$'':
\begin{equation}\label{IntroGammaInverse}\mbox{lim}_{z\to 0}e^{\phi(z)}\eta(z)\;\xi(0) e^{-\phi(0)} = 1\end{equation}
There is a 2-term $L_{\infty}$-action of the supersymmetry algebra on the large Hilbert space:
\begin{align}
d_{L_{\infty}}\;=\; & \underset{0}q + C^a\underset{1}q{}_a + C^a C^b\underset{2}q{}_{ab}\label{QRNS} \\
\underset{0}q\;=\; & Q'_{\rm RNS}\nonumber  \\
\underset{1}q{}_m\;=\; & \oint dz\;\partial X^m\nonumber  \\
\underset{1}q{}_{\alpha}\;=\; & \oint dz\; e^{-\phi/2}\Sigma_{\alpha}\nonumber  \\
\underset{2}q{}_{\alpha\beta}\;=\; & {1\over 2}\oint dz\; \Gamma^m_{\alpha\beta} \xi e^{-\phi}\psi^m\nonumber \end{align}
The verification of $d_{L_{\infty}}^2=0$ uses the identities:
\begin{align}
 & (e^{-\phi/2}\Sigma_{\alpha})(z)\;\;(e^{-\phi/2}\Sigma_{\beta})(w) \;=\;
{1\over z - w} \Gamma_{\alpha\beta}^m\left(\partial x^m + Q_{\rm RNS}'(\xi e^{-\phi}\psi^m)\right) + \mbox{regular}\nonumber  \\
 & (e^{-\phi}\psi^m)(z)\;\; (e^{-\phi/2}\Sigma_{\alpha})(w) \;=\;
{1\over z - w} \Gamma^m_{\alpha\beta} e^{-3\phi/2} \Sigma^{\beta}
+ \mbox{regular}\nonumber  \\
 & (\xi e^{-\phi}\psi^m)(z)\;\;(\xi e^{-\phi}\psi^n)(w) \;=\;
{1\over z - w}\delta^{mn} \;\xi\partial\xi\;e^{-2\phi}
+ \mbox{regular}\nonumber  \\
 & \Gamma^m_{\alpha\beta}\Gamma^m_{\gamma\delta} C^\alpha C^\beta C^\gamma = 0\nonumber \end{align}

\subsection{Strictification of supersymmetry in RNS formalism}\label{sec:IntroStrictification}

Let $G$ denote the supergroup generated by the supersymmetries
and $G_{\rm even}$ its even subgroup (generated by translations).

Following \cite{Berkovits:2021xwh}, we extend the RNS superstring by adding
auxiliary variables forming a ``BRST quartet''. The BRST quartet is
the chiral de Rham complex $\Omega^{\rm ch}_{G/G_{\rm even}}$,
a special case of a curved beta-gamma system, with the target space of the form $\Pi T(G/G_{\rm even})$.
The quartet fields are coordinates on $\Pi T(G/G_{\rm even})$,
called $\theta^{\alpha}$ and $\Lambda^{\alpha}$ in \cite{Berkovits:2021xwh},
and their conjugate momenta $p_{\alpha}$ and $\Omega_{\alpha}$. We take:
\begin{align}
V\;=\; & {\cal H}'_{RNS}\nonumber  \\
\widetilde{V}\;=\; & {\cal H}'_{RNS}\otimes \Omega^{\rm ch}_{G/G_{\rm even}}\nonumber \end{align}
There exists a $C$-dependent similarity transformation:
\begin{equation}R\;\in\;\mbox{End}_{C^{\infty}(\Pi \mathfrak{g})}
\left(
      C^{\infty}(\Pi \mathfrak{g})\otimes \widetilde{V}
      \right)\end{equation}
which brings the differential to the ``strict'' form, Eq. (\ref{IntroDStrict}):
\begin{align}
 & R\left(
       {1\over 2}C^aC^bf_{ab}^c{\partial\over\partial C^c} + d_{\rm quartet} +
       \right.\nonumber  \\
 & \left.
\quad
+ q_0
+ C^m \oint dz\; \partial X^m
+ C^{\alpha} \oint dz\; e^{-\phi/2}\Sigma_{\alpha}
+ {1\over 2}C^{\alpha}C^{\beta}\oint dz\; \Gamma^m_{\alpha\beta} \xi e^{-\phi}\psi^m
\right) R^{-1}\;=\nonumber  \\
=\; & C^aC^bf_{ab}^c{\partial\over\partial C^c} + \underset{0}{\widetilde{q}} + C^a\underset{1}{\widetilde{q}}{}_a \;=\nonumber  \\
=\; & d_0 + d_1\nonumber \end{align}
where
\begin{align}
 & d_0\;=\;R\left(
               {1\over 2}C^aC^bf_{ab}^c{\partial\over\partial C^c} + q_0
               + C^m \oint dz \partial X^m
               + C^{\alpha} \oint dz\; e^{-\phi/2}\Sigma_{\alpha}
               + {1\over 2}C^{\alpha}C^{\beta}\oint dz\; \Gamma^m_{\alpha\beta} \xi e^{-\phi}\psi^m
               \right) R^{-1}\nonumber  \\
 & d_1\;=\;Rd_{\rm quartet}R^{-1}\nonumber  \\
 & [R,C^a]\;=\;0\nonumber  \\
 & [d_0,C^a] \;=\; {1\over 2}f^a_{bc} C^bC^c\nonumber  \\
 & [d_1,C^a] \;=\; 0\nonumber  \\
 & I\;=\;p\circ R\circ j\nonumber \end{align}
where $p$ is the projection on the cohomology of $d_1$, and $j$ is the multiplication by
the unit operator of the quartet system:
\begin{align}
p\;:\; & \mbox{ker}(d_1)\rightarrow H(d_1)\nonumber  \\
j\;:\; & {\cal H}_{\rm RNS} \rightarrow {\cal H}_{\rm RNS}\otimes \Omega^{\rm ch}_{G/G_{\rm even}}\nonumber  \\
 & j\Phi = \Phi\otimes {\bf 1}\nonumber  \\
R\circ j\;:\; & C^{\infty}(\Pi\mathfrak{g})\otimes V
\rightarrow
C^{\infty}(\Pi\mathfrak{g})\otimes \widetilde{V}\nonumber \end{align}
The similarity transformation $R$ commutes with the multiplication by functions of the quartet
coordinate fields, and with the multiplication by functions of $C$.
(It depends on $C$ and on the auxiliary quartet coordinate fields, but does not contain neither derivatives
with respect to $C$, nor conjugate momenta to the quartet coordinates.)
Before the similarity transformation, the total complex is the free tensor product of two complexes:
\begin{align}
 & \left(
      C^{\infty}(\Pi\mathfrak{g})\otimes {\cal H}_{\rm RNS}
      \,,\;
      {1\over 2}C^aC^bf_{ab}^c{\partial\over\partial C^c} + q_0
      + \oint dz \left(
                       C^m \partial X^m
                       + C^{\alpha} e^{-\phi/2}\Sigma_{\alpha}
                       + {1\over 2}C^{\alpha}C^{\beta}\Gamma^m_{\alpha\beta} \xi e^{-\phi}\psi^m
                       \right)
      \right)
\otimes\nonumber  \\
 & \otimes
\left(
      \Omega^{\rm ch}_{G/G_{\rm even}}
      \;,\;
      d_{\rm quartet}
      \right)\nonumber \end{align}
Since $G/G_{\rm even}$ is contractible, $\Omega^{\rm ch}_{G/G_{\rm even}}$ is homotopy trivial
\cite{Malikov:1998dw}. Therefore the embedding
\begin{equation}\left(\,C^{\infty}(\Pi\mathfrak{g})\otimes {\cal H}_{\rm RNS}\,,\;d_{\rm RNS}\,\right)
\stackrel{\otimes 1}{\longrightarrow}
\left(\,
      C^{\infty}(\Pi\mathfrak{g})\otimes {\cal H}_{\rm RNS}\otimes \Omega^{\rm ch}_{G/G_{\rm even}}
      \,,\;
      d_{\rm RNS} + d_{\rm quartet}
      \,\right)\end{equation}
is a homotopy retract.

Therefore $C^{\infty}(\Pi\mathfrak{g})\otimes {\cal H}_{\rm RNS}$ is a homotopy retract of B-RNS-GSS:
\begin{equation}\left(\,
      C^{\infty}(\Pi\mathfrak{g})\otimes {\cal H}_{\rm RNS}
      \,,\;
      d_{\rm RNS}
      \,\right)
\stackrel{R\circ (\otimes 1)}{\longrightarrow}
\left(\,
      C^{\infty}(\Pi\mathfrak{g})\otimes {\cal H}_{\rm RNS}\otimes \Omega^{\rm ch}_{G/G_{\rm even}}
      \,,\;
      d_{\rm B-RNS-GSS}
      \,\right)\end{equation}
where
\begin{equation}d_{\rm B-RNS-GSS} = R (d_{\rm RNS} + d_{\rm quartet}) R^{-1}\end{equation}

\section{Finite-dimensional strictification}\label{Strictification}

Suppose that we have an $L_{\infty}$ action,
{\it i.e.} a collection of operators
$\underset{0}q$, $\underset{1}q{}_a$, $\underset{2}q{}_{ab}$, $\ldots$
acting on a linear space $V$, such that the operator:
\begin{align}
Q\;\in\; & \mbox{End}(C^{\infty}(\Pi \mathfrak{g})\otimes V)\nonumber  \\
d_{L_\infty}\;=\; & C_LC_L{\partial\over\partial C_L} + \underset{0}{q} + C_L^a \underset{1}q{}_a +
C_L^a C_L^b \underset{1}q{}_{ab} + \ldots\nonumber \end{align}
is nilpotent. Here $C_L$ denote the coordinates on $\Pi \mathfrak{g}$.
We added the index $L$, to distinguish them from the coordinates in the fiber of $\Pi TG$.
In fact, $u$ and $C_R$ will denote the coordinates on $\Pi TG$.
Let $G$ be the group manifold, $\mathfrak{g} = \mbox{Lie}(G)$,
and $C^a_R$ the coordinates on $\Pi \mathfrak{g}$.
Consider the operator $d_{\rm tot}$ acting on a larger space:
\begin{align}
d_{\rm tot}\;\in\; & \mbox{End}(C^{\infty}(\Pi\mathfrak{g})\otimes C^{\infty}(\Pi TG)\otimes V)\nonumber  \\
d_{\rm tot}\;=\; & {1\over 2} C_R^aC_R^b f_{ab}{}^c{\partial\over\partial C_R^c} +
C_R^a r_a \;+\nonumber  \\
 & + {1\over 2} C_L^aC_L^b f_{ab}{}^c{\partial\over\partial C_L^c} +
\underset{0}{q} +
C_L^a (\underset{1}q{}_a + l_a) + C_L^a C_L^b \underset{2}q{}_{ab} + \ldots\label{DefDtot}\end{align}
where $r_a$ are left-invariant vector fields on $G$ (infinitesimal {\bf right} shifts):
\begin{equation}\label{RightShift}(r_a \phi)(g) = - \left.{d\over  d\tau}\right|_{\tau=0} \phi(ge^{\tau t_a})\end{equation}
and $l_a$ are right-invariant vector fields on $G$:
\begin{equation}\label{LeftShifts}(l_a \phi)(g) = \left.{d\over  d\tau}\right|_{\tau=0} \phi(e^{\tau t_a}g)\end{equation}
We restrict ourselves to the formal neighborhood of the unit of $G$, where the group element
can be parametrized as an exponential of $u\in \mathfrak{g}$:
\begin{equation}g  = e^u\end{equation}
In Section \ref{GeneralAbsoluteSimilarityTransformation}
we will find the similarity transformation $R$ such that:
\begin{align}
R\;\in\; & \mbox{End}_{C^{\infty}(C_L,C_R,u)}\left(C^{\infty}(C_L,C_R,u)\otimes V\right)\nonumber  \\
Rd_{\rm tot}R^{-1}\;=\; & d_{\rm tot}'\label{DefRDtotRinv} \\
d_{\rm tot}'\;=\; & {1\over 2} C_R^aC_R^b f_{ab}{}^c{\partial\over\partial C_R^c} +
\underset{0}{q} +
C_R^a (\underset{1}q{}_a + r_a)  + C_R^a C_R^b \underset{2}q{}_{ab} + \ldots \;+\label{DTotPrime} \\
 & + {1\over 2} C_L^aC_L^b f_{ab}{}^c{\partial\over\partial C_L^c} +
C_L^a l_a\nonumber \end{align}
Thus, we obtain Eqs. (\ref{IntroDStrict}) and (\ref{IntroBicomplex}), where:
\begin{align}
d_{\rm strict}\;=\; & R d_{\rm tot} R^{-1}\nonumber  \\
d_0 \;=\; & R\left({1\over 2}C_R^a C_R^b f_{ab}{}^c{\partial\over\partial C_R^c} + C_R^a r_a\right) R^{-1}\nonumber  \\
d_1 \;=\; & R\left(
       {1\over 2} C_L^aC_L^b f_{ab}{}^c{\partial\over\partial C_L^c} +
       \underset{0}{q} +
       C_L^a (\underset{1}q{}_a + l_a) + C_L^a C_L^b \underset{2}q{}_{ab} + \ldots
       \right)R^{-1}\nonumber \end{align}
(Remember that we renamed $C$ to $C_L$.)

Notice that Eqs. (\ref{DefDtot}) and (\ref{DefRDtotRinv}) are related by $C_L \leftrightarrow C_R$
and $g\mapsto g^{-1}$.

An important for us special case is when a subalgebra $\mathfrak{h}\subset\mathfrak{g}$ acts strictly.
This means that if we restrict $C_L$ to $\Pi \mathfrak{h}\subset \Pi\mathfrak{g}$, the terms
$C_LC_L\underset{2}q$, $\ldots$ vanish. In this case it is possible to do a simpler (in some sense) similarity
transformation,
Section \ref{GeneralRelativeSimilarityTransformation}.
Then, there is not such relation between $d_{\rm tot}$ and $Rd_{\rm tot}R^{-1}$ as
$C_L \leftrightarrow C_R$ and $g\mapsto g^{-1}$.
In the field theory case,
Section \ref{ChiralStrictification},
$C_L$ and $C_R$ are very different, as $C_R$ is promoted to a chiral field while the ``background''
$C_L$ remains constant.

\section{Similarity transformation}\label{GeneralAbsoluteSimilarityTransformation}

We will interpret $C_L$, $C_R$, and $u$ as coordinates on the Q-manifold \cite{Alexandrov:1995kv}:
\begin{equation}\label{DefA}A = \frac{\Pi TG \times \Pi TG}{G}\end{equation}
Functions on $A$ are functions of $g_L\in G$, $g_R\in G$,
and their differentials, invariant under $(g_L,g_R)\mapsto (g_Lh, h^{-1}g_R)$.
Such functions can be constructed from the following  invariants:
\begin{align}
 & g = g_L g_R\nonumber  \\
 & C_L = d_Ag_L g_L^{-1}\nonumber  \\
 & C_R = - g_R^{-1}d_Ag_R\nonumber \end{align}
This establishes the identification  of functions of
$C_L$, $C_R$ and $u$ from
Section \ref{Strictification}
as functions on $A$,
and Eq. (\ref{DefDtot}) becomes:
\begin{align}
d_{\rm tot}\;\in\; & \mbox{End}(C^{\infty}(A)\otimes {\cal H})\nonumber  \\
d_{\rm tot}\;=\; & d_A +
\underset{0}{q} +
C_L^a \underset{1}q{}_a + C_L^a C_L^b \underset{2}q{}_{ab} + \ldots\nonumber \end{align}
Now it will become important that we work in the formal vicinity of the unit of $G$, which can be covered by the exponential map:
\begin{equation}g=e^u\,,\quad u\in \mathfrak g\end{equation}
Let us consider a larger $Q$-manifold
\begin{equation}\widetilde{A} \;=\; \Pi T {\mathbb R} \times A\end{equation}
and the following function on it:
\begin{align}
{\cal C}\;\in\; & C^{\infty}(\widetilde{A})\nonumber  \\
{\cal C} \;=\; & d_{\widetilde{A}}((g_L g_R)^{-t} g_L)g_L^{-1} (g_L g_R)^t =
-dt\, u + \widetilde{C}\nonumber \end{align}
where $t$ and $dt$ parametrize $\Pi T {\mathbb R}$. Explicitly:
\begin{align}
d_Ae^u e^{-u} \;=\; & {e^{{\rm ad}_u}-1\over {\rm ad}_u}du = C_L - e^{{\rm ad}_u}C_R\nonumber  \\
{\cal C}\;=\; & -dt u +
{e^{(1-t)\,{\rm ad}_u}-1\over e^{{\rm ad}_u} - 1}C_L +
{e^{-t\,{\rm ad}_u} - 1\over e^{-{\rm ad}_u} - 1}C_R\nonumber \end{align}

By construction, $\cal C$ satisfies:
\begin{align}
 & d_{\widetilde{A}}{\cal C} = {\cal C}^2\nonumber  \\
 & {\cal C}|_{t=0,dt=0} = C_L\nonumber  \\
 & {\cal C}|_{t=1,dt=0} = C_R\nonumber \end{align}
where $d_{\widetilde{A}}$ is the cohomological vector field on $\widetilde{A}$;
it includes $dt{\partial\over\partial t}$.
Consider the operators:
\begin{align}
{\cal Q}\;\in\; & \mbox{End}\left(C^{\infty}(\widetilde{A})\otimes {\cal H}\right)\nonumber  \\
{\cal Q}\;=\; & d_{\widetilde{A}} + \underset{0}{q} + {\cal C}^a \underset{1}{q}{}_a + {\cal C}^a {\cal C}^b \underset{2}{q}{}_{ab} + \ldots\nonumber \end{align}
To make contact with Eqs. (\ref{DefDtot}) and (\ref{DefRDtotRinv}), we observe:
\begin{align}
 & {\cal Q}^2 = 0\nonumber  \\
 & {\cal Q}|_{t=0, dt=0} = d_{\rm tot}\nonumber  \\
 & {\cal Q}|_{t=1, dt=0} = d_{\rm tot}'\nonumber \end{align}
Therefore the similarity transformation $R$ in Eq. (\ref{DefRDtotRinv}) can be obtained as a path ordered
exponential:
\begin{equation}\label{DefR}R = P\exp\left(\int_0^1 dt {\cal A}(t)\right)\end{equation}
where $\cal A$ is obtained as the coefficient of $dt$ in $\cal Q$:
\begin{align}
{\cal Q} \;=\; & {\cal Q}|_{dt=0} + dt \left({\partial \over \partial t} - {\cal A}(t)\right)\nonumber  \\
{\cal A}(t)\;=\; & u^a\left.{\partial\over\partial\zeta^a}\right|_{\zeta = {\cal C}(t,\;dt=0)}
\left( \zeta^a \underset{1}q{}_a + \zeta^a\zeta^b\underset{2}q{}_{ab} + \ldots\right)\label{DefCalA}\end{align}
Indeed, ${\cal Q}^2 = 0$ implies:
\begin{equation}{\partial\over\partial t}{\cal Q}_{dt=0} = [{\cal A},{\cal Q}_{dt=0}]\end{equation}
and therefore:
\begin{equation}{\cal Q}|_{t = t_0, dt = 0} = \left(P\exp\int_0^{t_0} dt [{\cal A}(t),\_]\right) {\cal Q}|_{t=0,dt=0}\end{equation}

\section{Partial (relative) similarity transformation}\label{GeneralRelativeSimilarityTransformation}

If a subalgebra $\mathfrak{h}\subset\mathfrak{g}$ acts strictly,
there is a more economical strictification procedure. The corresponding
``relative'' similarity transformation does not use generators of $\mathfrak{h}$, only the generators of
the complement of $\mathfrak{h}$ in $\mathfrak{g}$.
For $\mathfrak{g}$ supersymmetry algebra and $\mathfrak{h}$ translations this procedure was
introduced in \cite{Elliott:2022brv}. It is certainly more economical than the
general Bar constructions of \cite{KrizMay1995}, and slightly more economical
than using the full $\mathfrak{g}$ in
Section \ref{Strictification}.
In the case of RNS string, elements of
$\mathfrak{h}$ are translations and its complement are supersymmetries.
Therefore the similarity transformation is Taylor series in odd variables.
In this case the formal neighborhood of $H$ in $G$ is the whole $G$.

In our approach we use the similarity
transformation, Eq. (\ref{DefRDtotRinv}). This streamlines the homotopy transfer and allows to establish the
relation to \cite{Berkovits:2021xwh}. Use of similarity transformations is very common
in the pure spinor formalism.

\subsection{General case}\label{sec:GeneralRelative}

Given a subalgebra $\mathfrak{h}\subset \mathfrak{g}$, suppose that there is an $\mathfrak{h}$-invariant decomposition of linear spaces:
\begin{equation}\mathfrak{g} = \mathfrak{h} \oplus \mathfrak{g}_{\perp}\end{equation}
Suppose that the action of $\mathfrak{h}$ integrates to an action of the Lie supergroup $H$.
Consider the formal vicinity of $H\subset G$ in $G$. A group element can be represented as:
\begin{align}
g \;=\; & e^{v_{\perp}}g_{||}\label{DecompositionH} \\
g_{||}\;\in\; & H\nonumber  \\
v_{\perp}\;\in\; & \mathfrak{g}_{\perp}\nonumber \end{align}
Instead of $A$ we will consider:
\begin{equation}A_{\rm rel} = \frac{\Pi TG_L \times \Pi T(G_R/H)}{G} = \frac{\Pi TG_L \times \Pi TG_R}{G\times \Pi TH} = A / \Pi TH\end{equation}
Points of $A_{\rm rel}$ are pairs $(y_L, m_R)$ where $y_L\in \Pi TG_L$, $m_R\in \Pi T(G_R/H)$,
modulo the equivalence relation:
\begin{equation}(y_L,m_R) \simeq (y_L h, h^{-1}m_R)\end{equation}
for $h\in H$. As in
Section \ref{GeneralAbsoluteSimilarityTransformation},
we use the notations for functions on $A$:
\begin{align}
g = e^{v_{\perp}}g_{||}\;=\; & g_L g_R\nonumber  \\
C_L \;=\; & dg_Lg_L^{-1}\nonumber  \\
C_R \;=\; & -g_R^{-1}dg_R\nonumber \end{align}
Functions on $A_{\rm rel}$ are those functions on $A$ which are invariant under the right shift by $\Pi TH$.
The following $\Pi TH$-invariant functions can be choosen as coordinates on $A_{\rm rel}$:
\begin{equation}\label{CoordinatesOnARel}v_{\perp}\,,\quad C_L\,,\quad g_{||} C^{\perp}_R g_{||}^{-1}\end{equation}
Other functions on $A_{\rm rel}$ can be expressed in terms of them. In particular:
\begin{align}
dg_{||}g_{||}^{-1} + g_{||}C_R^{||}g_{||}^{-1}\;=\; & C_L^{||} + \ldots\label{GTViaC} \\
dv_{\perp} \;=\; & C_L^{\perp} - g_{||} C_R^{\perp} g_{||}^{-1} + \ldots\label{DVPerpViaC}\end{align}
where $\ldots$ stand for terms linear and higher order in $v_{\perp}$.
Eqs. (\ref{GTViaC}) and (\ref{DVPerpViaC}) follow from:
\begin{equation}dg_{||}g_{||}^{-1} + g_{||} C_R g_{||}^{-1}
=
e^{-v_{\perp}} C_L e^{v_{\perp}}  - e^{-v_{\perp}} de^{v_{\perp}}\end{equation}
Now we can define the interpolating ghost:
\begin{align}
{\cal C}\;\in\; & C^{\infty}(\Pi T{\mathbb R} \times A_{\rm rel})\nonumber  \\
{\cal C} \;=\; & d(e^{-tv_{\perp}}g_L)g_L^{-1}e^{tv_{\perp}}\;=\;\label{RelativeC} \\
=\; & d\left(e^{(1-t)v_{\perp}}g_{||} g_R^{-1}\right)g_R g_{||}^{-1} e^{-(1-t)v_{\perp}}\nonumber \end{align}
Therefore:
\begin{align}
{\cal C}_t \;=\; & -v_{\perp}\nonumber \end{align}
We define, as in Eq. (\ref{DefCalA}):
\begin{align}
{\cal A}(t)\;=\; & v^c_{\perp}\left.{\partial\over\partial\zeta^c}\right|_{\zeta = {\cal C}(t,\;dt=0)}
\left(\zeta^a \underset{1}q{}_a  + \zeta^a\zeta^b\underset{2}q{}_{ab} + \ldots\right)\label{CalA}\end{align}
Then ${\cal Q}^2 = 0$ implies:
\begin{equation}\label{TimeDerivativeOfQ}{\partial\over\partial t}{\cal Q}_{dt=0} = [{\cal A},{\cal Q}_{dt=0}]\end{equation}
We define $R$ as in Eq. (\ref{DefR}) with $\cal A$ given by Eq. (\ref{CalA}).
We denote a $C_{\rm rel}$ a combination of $C_L$ and $C_R$:
\begin{align}
C_{\rm rel} = {\cal C}|_{t=1,dt=0} \;=\; & dg_{||}g_{||}^{-1} + g_{||}C_R g_{||}^{-1}\label{RelativeC1}\end{align}
The result of the similarity transformation is:
\begin{align}
Rd_{\rm tot} R^{-1}\;=\; & d_A + \underset{0}{q} + C_{\rm rel}^a \underset{1}{q}{}_a + C_{\rm rel}^a C_{\rm rel}^b \underset{2}{q}{}_{ab} + \ldots\nonumber \end{align}
Suppose that for all $n>1$ and all $w\in\mathfrak{h}$:
\begin{equation}w^{a_1} \underset{n}q{}_{a_1 a_2\ldots a_n} = 0\end{equation}
In other words, the action of the subalgebra $\mathfrak{h}\subset\mathfrak{g}$ is strict.
Notice that the second term on the right hand side of Eq. (\ref{RelativeC1}) does not depend on $C_L$,
and the first term takes values in $\mathfrak{h}$.  Therefore:
\begin{equation}\label{CalCOnCL}{\partial\over\partial C_L} C_{\rm rel} \in\mathfrak{h}\end{equation}
Then, Eq. (\ref{CalCOnCL}) implies that
the dependence of $\widehat{Q}^R$ on $C_L$ is at most linear. In other words, this ``partial'' similarity transformation
is sufficient for obtaining the strict action.

\subsection{Special case when $\mathfrak{h}\subset \mathfrak{g}$ is center}\label{sec:SpecialRelative}

We are interested in the special case when $\mathfrak{h}$ is the center of $\mathfrak{g}$, and
\begin{equation}[\mathfrak{g}_{\perp},\mathfrak{g}_{\perp}]\subset \mathfrak{h}\end{equation}
In this case formulas simplify. We have:
\begin{equation}e^{u} = e^{v_{\perp}} g_{\parallel}\end{equation}
\begin{equation}   d e^{u} e^{-u}
=
C_L - e^{\mbox{ad}_u} C_R
=
d v_{\perp} + d g_{\parallel}  g_{\parallel}^{-1}
+
{1\over 2} [v_{\perp}, d v_{\perp}]\end{equation}
\begin{equation}d v_{\perp} = C_L^{\perp} - C_R^{\perp}\end{equation}
\begin{align}
d g_{\parallel}g_{\parallel}^{-1}\;=\; & C_L^{\parallel}
-
C_R^{\parallel}
-
[v_{\perp}, C_R^{\perp}]
-
\frac12 [v_{\perp}, d v_{\perp}]\nonumber  \\
=\; & C_L^{\parallel}
-
C_R^{\parallel}
-
{1\over 2} [v_{\perp}, C_L^{\perp} + C_R^{\perp}]\nonumber \end{align}
The coordinate functions defined in Eq. (\ref{CoordinatesOnARel}) are:
\begin{equation}v_{\perp}\,,\;C_L\,,\;C_R^{\perp}\end{equation}
The interpolating ghost is:
\begin{align}
{\cal C} \;=\; & - dt\, v_{\perp}
+
(1-t)\, d v_{\perp}
-
\frac{(1-t)^2}{2}[d v_{\perp}, v_{\perp}]\nonumber  \\
 & +
d g_{\parallel}g_{\parallel}^{-1}
+
C_R
+
(1-t)[v_{\perp}, C_R^{\perp}]\nonumber \end{align}

\begin{align}
{\cal C}^{\perp} \;=\; & - dt\, v_{\perp} + (1-t) C_L^{\perp} + t C_R^{\perp}\label{RelCPerp} \\
{\cal C}^{\parallel}\;=\; & C_L^{\parallel}
+
\left(
      \frac{t^2}{2} - t
      \right)
[v_{\perp}, C_L^{\perp}]
-
\frac{t^2}{2}
[v_{\perp}, C_R^{\perp}]\nonumber \end{align}
\begin{equation}C_{\rm rel} = C_R^{\perp} + C_L^{||} - {1\over 2} [v_{\perp},C_L^{\perp}+C_R^{\perp}]\end{equation}

\section{Strictification of space-time supersymmetry}\label{StrictificationOfRNS}

Let us now consider the case when $\mathfrak{g}$ is the ten-dimensional supersymmetry algebra,
and $V$ is the large Hilbert space ${\cal H}'_{\rm RNS}$.

\subsection{Similarity transformation}\label{sec:ZeroModeSimilarityTransformation}

Using the vector fields $\underset{1}{q}{}_a$ and $\underset{2}{q}{}_{ab}$ from Eq. (\ref{QRNS}),
Eq. (\ref{DefCalA}) gives:
\begin{align}
{\cal C}^{\alpha} \;=\; & -dt \theta^{\alpha} + (1-t)C_L^{\alpha} + t C_R^{\alpha}\nonumber  \\
{\cal C}^m \;=\; & - dt x^m
+ (1-t)C_L^m + {t(t-1)\over 2}(\theta^{\alpha}\Gamma^m_{\alpha\beta} C_L^{\beta})
+ t C_R^m - {t(t-1)\over 2}(\theta^{\alpha}\Gamma^m_{\alpha\beta} C_R^{\beta})\nonumber  \\
{\cal A}\;=\; & x^mP_m
+ \theta^{\alpha}e^{-\phi/2}\Sigma_{\alpha}
+ \theta^{\alpha} (tC_R^{\beta} + (1-t)C_L^{\beta}) \Gamma_{\alpha\beta}^m \xi e^{-\phi}\psi^m\nonumber \end{align}

\subsection{Relative version}\label{sec:ZeroModeRelative}

Taking $\mathfrak{h}\subset\mathfrak{g}$ the subalgebra of translations generated by $\oint \partial X$,
Section \ref{sec:SpecialRelative} gives:
\begin{align}
{\cal C}^{\alpha} \;=\; & -dt \theta^{\alpha} + (1-t)C_L^{\alpha} + t C_R^{\alpha}\nonumber  \\
{\cal C}^m \;=\; & C_L^m + \left({t^2\over 2} - t\right)(\theta^{\alpha}\Gamma^m_{\alpha\beta} C_L^{\beta})
- {t^2\over 2}(\theta^{\alpha}\Gamma^m_{\alpha\beta} C_R^{\beta})\nonumber  \\
C^{\alpha}_{\rm rel}\;=\; & C_R^{\alpha}\nonumber  \\
C^m_{\rm rel}\;=\; & C_L^m
- {1\over 2}\left(
                  (\theta^{\alpha}\Gamma^m_{\alpha\beta} C_L^{\beta})
                  +
                  (\theta^{\alpha}\Gamma^m_{\alpha\beta} C_R^{\beta})
                  \right)\nonumber  \\
{\cal A}\;=\; & \theta^{\alpha}e^{-\phi/2}\Sigma_{\alpha}
+ \theta^{\alpha} (tC_R^{\beta} + (1-t)C_L^{\beta}) \Gamma_{\alpha\beta}^m \xi e^{-\phi}\psi^m\nonumber \end{align}

\subsection{Why this is not the whole story}\label{sec:NotEnough}

However, for the purpose of relating to the pure spinor formalism, this is not enough.
It is not enough to simply pass from $V$ to a quasiisomorphic $\widetilde{V}$ with the strict action.
In fact there are various strictification procedures.
One has to choose one so that $\widetilde{V}$ is the Hilbert space of a well-defined CFT.
At this time, we do not have a general procedure for strictification of CFTs with $L_{\infty}$-action,
satisfying this property.
Straightforward application of Section \ref{sec:SpecialRelative} gave
us so far  a mixture of conformal fields ($X$, $\psi$, and RNS ghosts) and
the variables $C_R$ and $v_{\perp}$ which are constants on the worldsheet.
At least it is necessary to promote  $C_R$ and $v_{\perp}$ from constants to chiral fields.

In the rest of this paper we will describe an {\it ad hoc} procedure working
in the context of \cite{Berkovits:2021xwh}. Strictly speaking, it does not follow from
a functorial procedure. But it is very close to the finite-dimensional strictification
procedure of
Section \ref{Strictification} and
Section \ref{sec:SpecialRelative}.
It is almost enough to just allow non-constant  $C_R$,
but with corrections which we will describe in the next sections.
It is important that the BRST differential
$\underset{0}q$ of the RNS string is of the Faddeev-Popov type, {\it i.e.} obtained by gauge-fixing worldsheet
super-diffeomorphisms.

\section{Chiral analogue of strictification}\label{ChiralStrictification}

In the chiral case the finite-dimensional supermanifold $\Pi TG$ parametrized by $C_R$ and $u$
has to be replaced with the chiral de Rham complex, which we will now briefly review.

\subsection{Chiral de Rham complex}\label{sec:ChiralDeRham}

For a complex manifold  satisfying certain conditions,
one can associate a conformal field theory known as ``curved $\beta\gamma$-system''
\cite{Malikov:1998dw} \cite{Nekrasov:2005wg}.
Its BRST complex is called ``chiral de Rham complex''. In particular, one can always associate a
chiral de Rham complex to the odd tangent bundle $\Pi TM$ to any complex supermanifold $M$, which we will
denote $\Omega^{\rm ch}_M$. In particular, for our symmetry group $G$,
there is the chiral de Rham complex $\Omega^{\rm ch}_G$.
We will work in the formal vicinity of the unit of $G$;
let $Y(z)$ denote the chiral operators corresponding to the coordinates on $\Pi TG$,
and $P(z)$ the corresponding momenta.
There are two types of dimension one operators in $\Omega_G^{\rm ch}$. One is corresponding to
holomorphic vector fields  $V\in \mbox{Vect}(\Pi TG)$:
\begin{equation}\label{VectorType}V^i(Y(z))\; P_i(z)\end{equation}
and another is corresponding
to holomorphic one-forms on $B\in \Omega^1(\Pi TG)$:
\begin{equation}\label{OneFormType}B_i(Y(z))\; {\partial\over\partial z}Y^i(z)\end{equation}
We call them "vector-type" and "one-form-type",
correspondingly. Generally speaking, vector-type currents mix into one-form-type currents.

The canonical odd nilpotent vector field on $\Pi TG$ is a symmetry of $\Omega_G^{\rm ch}$,
and the corresponding conserved current will be denoted $\mathfrak{d}$:
\begin{equation}d_{\Omega_G^{\rm ch}} = \oint \mathfrak{d}\end{equation}

\subsection{CFTs with $L_{\infty}$-action of $\mathfrak{g}$}\label{sec:CFTLinfty}

Suppose that we have a conformal field theory, with the space of states $\cal H$.
We identify the space of states with the space of local operators.
Suppose that there is an $L_{\infty}$ action of a finite-dimensional
Lie superalgebra $\mathfrak{g}$ on $\cal H$, {\it i.e.} Eq. (\ref{IntroLinfty}) with
$V = {\cal H}$ and $\underset{n}q$ generated by local currents $\underset{n}j$. In other words:
\begin{align}
Q\;\in\; & \mbox{End}(C^{\infty}(\Pi \mathfrak{g})\otimes {\cal H})\nonumber  \\
Q\;=\; & {1\over 2} C^aC^b f_{ab}{}^c{\partial\over\partial C^c} +
\oint \underset{0}j +
C^a \oint \underset{1}j{}_a + C^a C^b \oint \underset{2}j{}_{ab} + \ldots\label{LInftyQCFT} \\
Q^2=0 & \nonumber \end{align}

\subsection{Chiral de Rham as a BRST quartet}\label{sec:ChiralDeRhamQuartet}

We must stress that  $C^a$ of Eq. (\ref{LInftyQCFT}) are just auxiliary Grassmann variables, not quantum fields.

We want to consider the chiral analogue of the Q-manifold $A$ defined in Eq. (\ref{DefA}).
For clarity, we will distinguis the two copies of $\Pi TG$ by the indices $L$ and $R$.
We consider the linear superspace:
\begin{equation}C^{\rm ch} = \left(C^{\infty}(\Pi TG_L)\;\otimes \; \Omega^{\rm ch}_{G_R}\right)^G\;\simeq\; C^{\infty}(\Pi\mathfrak{g})\otimes \Omega^{\rm ch}_{G_R}\end{equation}
Here $G$ acts by right shifts on $G_L$ and by left shifts on $G_R$.
The coordinates on $\Pi\mathfrak{g}$ are denoted $C_L^a$.

Any function $f\in C^{\infty}(\Pi TG_R)$ defines a chiral operator $f(g)$ in $\Omega^{\rm ch}_{G_R}$.
Then:
\begin{equation}f(\,g_L\;\; g_R(z)\,) \in C^{\rm ch}\end{equation}
Here $g_L$ is constant and $g_R(z)$ is holomorphic. Consider the following odd nilpotent operator:
\begin{align}
Q^{\rm tot} \;=\; & \mbox{End}\left(
                (C^{\infty}(\Pi TG_L)\otimes \Omega^{\rm ch}_{G_R})^G\otimes {\cal H}
                \right)\nonumber  \\
Q^{\rm tot} \;=\; & d_{\Pi TG_L} + d_{\Omega^{\rm ch}_{G_R}} + \oint \underset{0}j +
C_L^a \oint \underset{1}j{}_a + C_L^a C_L^b \oint \underset{2}j{}_{ab} + \ldots\label{ChiralQtot}\end{align}
This defines an $L_{\infty}$-action of $\mathfrak{g}$ on $\Omega^{\rm ch}_G\otimes {\cal H}$.

\subsection{Some  holomorphic currents}\label{sec:ChiralSimTrans}

The space
$(C^{\infty}(\Pi TG_L)\otimes \Omega^{\rm ch}_{G_R})^G \simeq C^{\infty}(\Pi\mathfrak{g})\otimes \Omega^{\rm ch}_{G_R}$
consists of local operators of $\Omega^{\rm ch}_{G_R}$ depending parametrically on the background ghosts $C_L$.
There is a subspace:
\begin{equation}C^{\infty}\left({\Pi TG_L \times \Pi TG_R\over G}\right)
\subset
(C^{\infty}(\Pi TG_L)\otimes \Omega^{\rm ch}_{G_R})^G\end{equation}
consisting of those operators which are built only out of $Y^i(z)$
(no dependence on the conjugate momenta $P_i(z)$).

Suppose that the $L_{\infty}$-action of $\mathfrak{g}$ on $\cal H$ is generated by holomorphic
currents, {\it i.e.} operators of conformal dimension $(1,0)$. We will denote the space
of holomorphic currents:
\begin{equation}{\cal H}_{(1,0)} \subset {\cal H}\end{equation}
Consider the conserved current:
\begin{align}
{\cal J}\;\in\; & C^{\infty}\left({\Pi TG_L \times \Pi TG_R\over G}\right) \otimes {\cal H}_{(1,0)}\nonumber  \\
{\cal J}\;=\; & \left.
\left(
      {\cal C}^a(t) \underset{1}j{}_a +
      {\cal C}^a(t){\cal C}^b(t)\underset{2}j{}_{ab} + \ldots
      \right)
\right|_{dt=0}\nonumber  \\
{\cal A}(t)\;=\; & v^c_{\perp}\left.{\partial\over\partial\zeta^c}\right|_{\zeta = {\cal C}(t,\;dt=0)}
\left(\underset{1}j{}_a \zeta^a + \underset{2}j{}_{ab}\zeta^a\zeta^b + \ldots\right)\nonumber \end{align}
where ${\cal C}(t)$ is given by the same Eq. (\ref{RelativeC}), except now $g_R$ is not a constant element of $G$, but a
holomorphic function of $z$. The naive analogue of Eq. (\ref{TimeDerivativeOfQ}) would be:
\begin{equation}\label{TentativeDt}{\partial\over\partial t} \oint {\cal J}(t)
=
\left[
      \left(
            d_{\Pi TG_L} + d_{\Omega^{\rm ch}_{G_R}} +
            \oint {\cal J}(t)
            \right),
      \oint {\cal A}(t)
      \right]
\quad \mbox{(wrong)}\end{equation}
However, Eq. (\ref{TentativeDt}) is not true, because the currents
$\underset{p}j{}_{a_1\ldots a_p}$ do not form an $L_{\infty}$-algebra in any obvious sense.
Only their contour integrals (charges)
generate an $L_{\infty}$-algebra. Since $g_R$ is a non-constant function of $z$,
${\cal C}(t)$ is not a constant,
and $\oint {\cal J}$ can not be expressed in terms of $\oint \underset{p}j{}_{a_1\ldots a_p}$.
However, Eq. (\ref{TentativeDt}) holds up to terms containing the derivatives of $Y(z)$:
\begin{equation}\label{DtModDerivatives}{\partial\over\partial t} \oint {\cal J}(t)
=
\left[
      \left(
            d_{\Pi TG_L} + d_{\Omega^{\rm ch}_{G_R}} +
            \oint {\cal J}(t)
            \right),
      \oint {\cal A}(t)
      \right]
\quad \mbox{mod}\;\;\partial Y\end{equation}
Since ${\cal A}(t)$ does not contain the conjugate momenta $P_i$ of $Y^i$, the space of operators
containing $\partial Y$ is closed under $[{\cal A}(t),\_]$. This implies that:
\begin{align}
 & R\left(
       d_{\Pi TG_L} + d_{\Omega^{\rm ch}_{G_R}} +
       \oint \underset{0}j + C_L^a \oint \underset{1}j{}_a + C_L^a C_L^b \oint\underset{2}j{}_{ab} + \ldots
       \right)R^{-1}\;=\nonumber  \\
=\; & d_{\Pi TG_L} + d_{\Omega^{\rm ch}_{G_R}} +
\oint \underset{0}j +
\oint C_R^a \underset{1}j{}_a +  \oint C_R^a C_R^b\underset{2}j{}_{ab} + \ldots
\quad \mbox{mod}\;\;\partial Y\nonumber \end{align}
This has strict dependence on $C_L$ ({\it via} $d_{\Pi TG_L}$), modulo terms which contain $\partial Y$,
{\it i.e.} $\partial u$ and $\partial C_R$. Therefore, the naive chiral strictification procedure
works only up to one-form-type operators of $\Omega^{\rm ch}_{G_R}$.
These one-form-type operators need to be studied on case-by-case basis.
If they are constant-linear function of $C_L$, then the strictification is achieved.
One can also modify $R$ by $\partial Y$-dependent terms.

In the next section we will see how this works in the case of RNS superstring.

\section{B-RNS-GSS}\label{SigmaModelSimilarityTransformation}

We will now show that the strictification procedure of
Section \ref{GeneralRelativeSimilarityTransformation},
corrected by some one-form-type operators, works for the RNS string.

\subsection{Similarity transformation from Weil to Cartan model}\label{WorldsheetSUSY}

The chiral analogue of the procedure of
Section \ref{GeneralRelativeSimilarityTransformation}
is not completely straightforward.
Following \cite{Berkovits:2021xwh}
we have to first do another similarity transformation, which involves local operators
proportional to derivatives $\partial Y$ (the one-form-type operators).
This is a general theme in curved beta-gamma systems \cite{Nekrasov:2005wg}.
The center-of-mass equations can be canonically lifted to the vector-type operators
of the form Eq. (\ref{OneFormType}), but this is not enough. One has to also add appropriate one-form-type
operators, Eq. (\ref{VectorType}). We will now describe this additional similarity transformation, which has to be done first.

\subsubsection{Another example of $L_{\infty}$-action}\label{sec:GeneralExampleLInfty}

By a coincidence, we will need yet another geometrical construction involving an $L_{\infty}$-action,
which we will now describe.

Consider a Lie algebra $\mathfrak a$ and a subalgebra $\mathfrak{b}\subset\mathfrak{a}$, with an
$\mathfrak{b}$-invariant decomposition:
\begin{equation}\mathfrak{a} = \mathfrak{b} \oplus \mathfrak{a}_{\perp}\end{equation}
Let $U$ be a linear space with an action of $\mathfrak{b}$ ({\it i.e.} a $\mathfrak{b}$-module).
There is no general construction to extend an action of $\mathfrak{b}$ to an action of $\mathfrak{a}$.
But if $U$ is a {\bf differential} $\mathfrak{b}$-module, {\it i.e.} for all $\xi\in \mathfrak{b}$
are defined operators ${\cal L}_{\xi}$ and $\iota_{\xi}$ satisfying:
\begin{equation}[\iota_{\xi},d] = {\cal L}_{\xi}\end{equation}
then the action of $\mathfrak{b}$ can be extended to a two-term $L_{\infty}$-action of $\mathfrak{a}$.
The construction goes as follows. Let $c^a$ denote the coordinates on $\Pi \mathfrak{b}$,
and $\gamma^{\alpha}$ coordinates on $\Pi \mathfrak{a}_{\perp}$.
The BRST operator of $\mathfrak{a}$ with trivial coefficients is:
\begin{align}
d^0_{\mathfrak{a}}c^a \;=\; & {1\over 2} f^a_{bc} c^b c^c + {1\over 2} f^a_{\alpha\beta} \gamma^{\alpha}\gamma^{\beta}\nonumber  \\
d^0_{\mathfrak{a}}\gamma^{\alpha} \;=\; & f^{\alpha}_{a\beta} c^a\gamma^{\beta}\nonumber \end{align}
The following map provides a homomorphism of the Weil algebra of $\mathfrak{b}$ into
$C^{\infty}(\Pi \mathfrak{a})$:
\begin{align}
{\bf a}^a \;=\; & c^a\nonumber  \\
{\bf f}^a \;=\; & {1\over 2} f^a_{\alpha\beta} \gamma^{\alpha}\gamma^{\beta}\nonumber \end{align}
The two-term $L_{\infty}$-action of $\mathfrak{a}$ on $U$ is given by the nilpotent operator:
\begin{align}
e^{-\iota^U_{\bf a}}(d^0_{\mathfrak{a}} + d_U)e^{\iota^U_{\bf a}}\;\in\; & \mbox{End}(C^{\infty}(\Pi\mathfrak{a})\otimes U)\nonumber  \\
e^{-\iota^U_{\bf a}}(d^0_{\mathfrak{a}} + d_U)e^{\iota^U_{\bf a}}\;=\; & d^0_{\mathfrak{a}} + d_U
+ c^a l^U_a
+ {1\over 2}f^a_{\alpha\beta} \gamma^{\alpha}\gamma^{\beta} \iota^U_a\nonumber \end{align}
This is a variation on the usual relation between Weil and BRST models of equivariant
cohomology \cite{Cordes:1994fc}, \cite{Meinrenken}.
Here we need a slight generalization of this construction.
Let $V$ be an $\mathfrak{a}$-module, and $d_{\mathfrak{a}}^V\in \mbox{End}(C^{\infty}(\Pi\mathfrak{a})\otimes V)$
the corresponding nilpotent operator. Then, consider the two-term $L_{\infty}$-action of $\mathfrak{a}$ on
$V\otimes U$, with the following differential:
\begin{align}
e^{-\iota^U_{\bf a}}(d^V_{\mathfrak{a}} + d_U)e^{\iota^U_{\bf a}}\;\in\; & \mbox{End}(C^{\infty}(\Pi\mathfrak{a})\otimes V\otimes U)\nonumber  \\
e^{-\iota^U_{\bf a}}(d^V_{\mathfrak{a}} + d_U)e^{\iota^U_{\bf a}}\;=\; & d^0_{\mathfrak{a}}
+ d_U
+ \gamma^{\alpha} l^V_{\alpha}
+ c^a (l^V_a + l^U_a)
+ {1\over 2}f^a_{\alpha\beta} \gamma^{\alpha}\gamma^{\beta} \iota^U_a\nonumber \end{align}

\subsubsection{Special case of RNS superstring}\label{sec:WeilCartanRNS}

In the particular case of \cite{Berkovits:2021xwh} we need:
\begin{align}
\mathfrak{a}\;=\; & \mbox{SVir}\nonumber  \\
\mathfrak{b}\;=\; & \mbox{Vir}\nonumber  \\
V\;=\; & {\cal H}_{X,\psi}\nonumber  \\
U\;=\; & \Omega^{\rm ch}_{G/G_{\rm even}}\nonumber \end{align}
where $\mbox{Vir}$ is the Virasoro algebra (worldsheet conformal transformations) and $\mbox{SVir}$ the
super-Virasoro algebra of the superconformal transformations of the RNS worldsheet.
The BRST operator of the RNS string comes from the Faddeev-Popov gauge fixing of the diffeomorphisms of
the super-worldsheet.
In terms of the bosonic worldsheet $\Sigma$, the ghosts $c^z$ and $c^{\bar{z}}$ are coordinates on $\Pi \mbox{Vect}\Sigma$,
while $\gamma$ and $\overline{\gamma}$ are the ghosts of super-reparametrizations.

It is straightforward to generalize the finite-dimensional similarity transformation $R$ to the chiral case
so that it is $\rm Vir$-base. This means that it:

\begin{itemize}
\item commutes with
$f_{ab}^c c^b{\partial\over\partial c^c} + f_{a\alpha}^{\beta}\gamma^{\alpha}{\partial\over\partial\gamma^{\beta}} + l_a^V + l_a^U$ (in other words, it is conformally invariant)
\item does not involve $c^a$

\end{itemize}

Therefore, to compute the effect of the similarity transformation, it is enough to consider:
\begin{align}
\left.e^{-\iota^U_{\bf a}}(d^V_{\mathfrak{a}} + d_U)e^{\iota^U_{\bf a}}\right|_{\rm Vir-base}\;=\; & \oint \left(
            \gamma^2 b
            + \mathfrak{d}
            + \gamma \psi^m\partial X^m
            + \gamma^2 \partial \theta^{\alpha}\Omega_{\alpha}
            \right)\nonumber \end{align}
Here $\gamma^2 b$ comes from $d^0_{\mathfrak{a}}$, $\mathfrak{d}$ is the density of the
chiral de Rham differential $d_U = d_{\Omega^{\rm ch}_{G/G_{\rm even}}}$, the term
$\gamma\psi^m\partial X^m$ is the worldsheet supersymmetry with the ghost parameter $\gamma$,
and $\gamma^2\partial\theta^{\alpha} \Omega_{\alpha}$ is the curvature part $\iota_{\bf f}$
of the Cartan differential.
The full differential $Q^{\rm tot}$ of Eq. (\ref{ChiralQtot}) contains also the piece:
\begin{equation}d_{\Pi TG_L} + C_L^a \oint \underset{1}j{}_a + C_L^a C_L^b \oint \underset{2}j{}_{ab}\end{equation}
which in our case commutes with $\iota^U_{\bf a}$, and therefore carries over unchanged
by the similarity transformation with $e^{\iota^U_{\bf a}}$.

\subsection{Chiral analogue of the relative similarity transformation}\label{sec:ChiralRelative}

We will use the relative similarity transformation of
Section \ref{GeneralRelativeSimilarityTransformation}.
We take $\mathfrak{h}$ to be the algebra of translations, and $\mathfrak{g}_{\perp}$ supersymmetries:
\begin{equation}{\rm susy} = \mathfrak{h}\oplus \mathfrak{g}_{\perp}\end{equation}
as a linear space.

Using the notations of \cite{Berkovits:2021xwh}, $C_R^{\alpha}$ becomes chiral field $\Lambda^{\alpha}$,
and $v_{\perp}^{\alpha}$ becomes $\theta^{\alpha}$:
\begin{align}
\Lambda^{\alpha} \;=\; & C_R^{\alpha}\nonumber  \\
\theta^{\alpha} \;=\; & v_{\perp}^{\alpha}\nonumber \end{align}
The conjugate momenta of $\theta^{\alpha}$ and $\Lambda^{\alpha}$ are called $p_{\alpha}$ and
$\Omega_{\alpha}$.
The worldsheet action of $\Omega^{\rm ch}_{G_R/H}$ is:
\begin{equation}\label{QuartetAction}\int d^2z \;\left(p_{\alpha}\overline{\partial}\theta^{\alpha}  + \Omega_{\alpha}\overline{\partial}\Lambda^{\alpha}\right)\end{equation}
The de Rham density is:
\begin{equation}\mathfrak{d} = - \Lambda^{\alpha} p_{\alpha}\end{equation}
(The minus sign is because we use the right action, see Eq. (\ref{RightShift}).)
The formulas of Section \ref{sec:ZeroModeRelative} become:
\begin{align}
{\cal C}^{\alpha}(t) \;=\; & -dt \theta^{\alpha} + (1-t)C_L^{\alpha} + t \Lambda^{\alpha}\nonumber  \\
{\cal C}^m(t) \;=\; & C_L^m
+ \left( {t^2\over 2} - t \right)\Gamma_{\alpha\beta}^m \theta^{\alpha}C_L^{\beta}
- {t^2\over 2}\Gamma_{\alpha\beta}^m \theta^{\alpha}\Lambda^{\beta}\nonumber  \\
{\cal J}(t)\;=\; & {\cal C}^m(t)\partial X^m +
{\cal C}^{\alpha}(t)e^{-\phi/2}\Sigma_{\alpha} +
{1\over 2}{\cal C}^{\alpha}(t){\cal C}^{\beta}(t)\Gamma_{\alpha\beta}^m \xi e^{-\phi}\psi^m\nonumber  \\
{\cal A}(t)\;=\; & \theta^{\alpha}e^{-\phi/2}\Sigma_{\alpha}
+ \theta^{\alpha} (t\Lambda^{\beta} + (1-t)C_L^{\beta}) \Gamma_{\alpha\beta}^m \xi e^{-\phi}\psi^m\nonumber \end{align}
We have:
\begin{align}
\iota_{\bf a}\;=\; & \oint c\partial\theta^{\alpha} \Omega_{\alpha}\nonumber  \\
\iota_{\bf f}\;=\; & \oint \gamma^2\partial\theta^{\alpha} \Omega_{\alpha}\nonumber \end{align}
Importantly, $\cal A$ is $\mbox{Vir}$-base,
{\it i.e.} conformally invariant and does not contain $c$. The similarity
transformation $e^{\iota_{\bf a}}$ introduced $c^a l^U_a$ into the
differential. In other words, it ``extended the conformal transformations to $\Lambda^{\alpha}$
and $\theta^{\alpha}$''. Without this, expressions like $\oint \theta^{\alpha}e^{-\phi/2}\Sigma_{\alpha}$
would not be conformally invariant, complicating the commutators. But due to the $e^{\iota_{\bf a}}$ similarity
transformation, we have a relatively simple form of the $\partial Y$-terms:
\begin{align}
 & \left[\;
      e^{\iota_{\bf a}}
      \left(
            d_{\Pi TG_L} + d_{\Omega^{\rm ch}_{G_R/H}} +
            \oint {\cal J}(t)
            \right)
      e^{-\iota_{\bf a}}
      \,,\;
      \oint {\cal A}(t)
      \;
      \right]\;=\nonumber  \\
=\; & {\partial\over\partial t} \left(e^{\iota_{\bf a}}\oint {\cal J}(t)e^{-\iota_{\bf a}}\right)
+
\left[\;
      \iota_{\bf f}
      \,,\;
      \oint {\cal A}(t)
      \;
      \right]\label{ExtraTerm}\end{align}
The use of relative similarity transformation (as opposed to the absolute one) simplified the
computation. First, because the OPEs of the currents in $\cal A$ have only first order poles
with $\cal J$. If we have used the absolute similarity transformation
of Section \ref{GeneralAbsoluteSimilarityTransformation},
the presence of the
double pole in the OPE $\partial X(z) \partial X(0)$ would have complicated the computation, leading to
more terms with derivatives $\partial Y$. Second, $d_{\Pi TG_L}$ involves the supersymmetry transformation,
and $\partial X$ is supersymmety invariant only up to total derivative. Had we used the absolute similarity
transformation, we would have needed to act by supersymmetry on $\oint C_R^m\partial X^m$, again generating
$\partial Y$-type terms (those containing $\partial C_R$).

Eq. (\ref{ExtraTerm}) implies that the only difference between the finite-dimensional Eq. (\ref{TimeDerivativeOfQ})
and its chiral analogue Eq. (\ref{ExtraTerm}) is due to the
curvature term $\iota_{\bf f}$ in the Cartan differential:
\begin{align}
\left[\iota_{\bf f},\oint {\cal A}(t)\right]\;= & \left[\;
      \oint \gamma^2\;\partial \theta^{\alpha}\;\Omega_{\alpha}
      \,,\;
      \oint {\cal A}(t)
      \;
      \right]
\;=\;\nonumber  \\
=\; & \left[\;
      \oint \gamma^2\;\partial \theta^{\alpha}\;\Omega_{\alpha}
      \,,\;
      \oint
      \theta^{\alpha} (t\Lambda^{\beta} + (1-t)C_L^{\beta}) \Gamma_{\alpha\beta}^m \xi e^{-\phi}\psi^m
      \;
      \right]\;=\nonumber  \\
=\; & t \oint \gamma \psi^m \theta^{\alpha}\Gamma_{\alpha\beta}^m\partial\theta^{\beta}\nonumber \end{align}
(We have used Eq. (\ref{IntroGammaInverse}).) Then,
$\left[\left[\;
            \iota_{\bf f}
            \,,\;
            \oint{\cal A}(t_1)\;\right]\;\oint{\cal A}(t_2)\right]$ is:
\begin{equation}\left[\;
      \oint t_1\gamma \psi^m \theta^{\alpha}\Gamma_{\alpha\beta}^m\partial\theta^{\beta}
      \,,\;
      \oint {\cal A}(t_2)
      \;
      \right]
\;=\;
- t_1\oint
\theta^{\alpha}\Gamma_{\alpha\beta}^m\partial\theta^{\beta}
\;
\theta^{\gamma}(t_2\Lambda^{\delta} + (1-t_2)C_L^{\delta}) \Gamma_{\gamma\delta}^m\end{equation}
and $\left[\left[\left[\;
                  \iota_{\bf f}
                  \,,\;
                  \oint{\cal A}(t_1)\;
                  \right]
                 \;\oint{\cal A}(t_2)\right]
           \;\oint{\cal A}(t_3)\right]$
is zero:
\begin{equation}\left[\;
      \oint
      \theta^{\alpha}\Gamma_{\alpha\beta}^m\partial\theta^{\beta}
      \;
      \theta^{\gamma}(t\Lambda^{\delta} + (1-t)C_L^{\delta}) \Gamma_{\gamma\delta}^m
      \,,\;
      \oint {\cal A}(t)
      \;
      \right]
\;=\;0\end{equation}
As in the finite-dimensional case, we denote:
\begin{align}
R\;=\; & P\exp\int_0^1 dt\oint {\cal A}(t)\nonumber \end{align}
This is a local holomorphic symmetry transformation. ``Local'' means mapping local operators into local operators.
Indeed, RNS  currents $e^{-\phi/2}\Sigma_{\alpha}$ and $\psi^m e^{-\phi}\xi$ are holomorphic,
and $\Lambda^{\alpha}$ and $\theta^{\alpha}$ are chiral scalar field.
Therefore, ${\cal H}'_{\rm RNS}\otimes \Omega^{\rm ch}_{G/G_{\rm even}}$ has infinitely many local holomorphic
symmetries, and $R$ is one of them.
\begin{align}
 & Re^{\iota_{\bf a}}
\left(
      d_{\Pi TG_L} + d_{\Omega^{\rm ch}_{G_R/H}} +
      C_L^a \oint \underset{1}j{}_a + C_L^a C_L^b \oint \underset{2}j{}_{ab} + \ldots
      \right)
e^{-\iota_{\bf a}}R^{-1}\;=\nonumber  \\
=\; & d_{\Pi TG_L} + d_{\Omega^{\rm ch}_{G_R/H}} +\nonumber  \\
 & + \oint \Lambda^{\alpha} \Sigma_{\alpha} e^{-\phi/2}
+ {1\over 2}\oint \Lambda^{\alpha}\Lambda^{\beta}\Gamma_{\alpha\beta}^m \psi^m e^{-\phi} \xi
+ \oint \left(
              C_L^m
              - {1\over 2} \Gamma_{\alpha\beta}^m\theta^{\alpha}(C_L^{\beta}+\Lambda^{\beta})
              \right)\partial X^m
\;+\nonumber  \\
 & + \int_0^1 dt \;t \oint \gamma\psi^m\theta^{\alpha}\Gamma_{\alpha\beta}^m\partial\theta^{\beta}
- \int_0^1 dt_1 \int_{t_1}^1 dt_2 \;t_1
\theta^{\alpha}\Gamma_{\alpha\beta}^m\partial\theta^{\beta}
\;
\theta^{\gamma}(t_2\Lambda^{\delta} + (1-t_2)C_L^{\delta}) \Gamma_{\gamma\delta}^m
\;=\nonumber  \\
=\; & d_{\Pi TG_L} + d_{\Omega^{\rm ch}_{G_R/H}} +\nonumber  \\
 & + \oint \Lambda^{\alpha} \Sigma_{\alpha} e^{-\phi/2}
+ {1\over 2}\oint \Lambda^{\alpha}\Lambda^{\beta}\Gamma_{\alpha\beta}^m \psi^m e^{-\phi} \xi
+ \oint \left(
              C_L^m
              - {1\over 2} \Gamma_{\alpha\beta}^m\theta^{\alpha}(C_L^{\beta}+\Lambda^{\beta})
              \right)\partial X^m
\;+\nonumber  \\
 & + \oint
\left(
      {1\over 2} \gamma\psi^m\theta^{\alpha}\Gamma_{\alpha\beta}^m\partial\theta^{\beta}
      - {1\over 8}
      \theta^{\alpha}\Gamma_{\alpha\beta}^m\partial\theta^{\beta}
      \;
      \theta^{\gamma}\Gamma_{\gamma\delta}^m \Lambda^{\delta}
      - {1\over 24}
      \theta^{\alpha}\Gamma_{\alpha\beta}^m\partial\theta^{\beta}
      \;
      \theta^{\gamma}\Gamma_{\gamma\delta}^m C_L^{\delta}
      \right)\nonumber \end{align}
We re-group the terms to get:
\begin{align}
 & Re^{\iota_{\bf a}}
\left(
      d_{\Pi TG_L} + d_{\Omega^{\rm ch}_{G_R/H}} +
      \oint \underset{0}j +
      C_L^a \oint \underset{1}j{}_a + C_L^a C_L^b \oint \underset{2}j{}_{ab} + \ldots
      \right)
e^{-\iota_{\bf a}}R^{-1}\;=\nonumber  \\
=\; & \oint\left(
             cT + c\partial c b +
             \gamma \left(
                          \partial X^m +
                          {1\over 2}\theta^{\alpha}\Gamma_{\alpha\beta}^m\partial\theta^{\beta}
                          \right)\psi^m +
             \gamma^2 \partial\theta^{\alpha}\Omega_{\alpha}
             \right)\;+\nonumber  \\
 & +\;\oint \Lambda^{\alpha}\left(
                               \Sigma_{\alpha} e^{-\phi/2}
                               - p_{\alpha}
                               - {1\over 2} \partial X^m\Gamma^m_{\alpha\beta}\theta^{\beta}
                               - {1\over 8}
                               \Gamma^m_{\alpha\beta}\theta^{\beta}
                               \theta^{\gamma}\Gamma_{\gamma\delta}^m\partial\theta^{\delta}
                               \right)\;+\nonumber  \\
 & + \oint {1\over 2}\Lambda^{\alpha}\Lambda^{\beta}\Gamma_{\alpha\beta}^m \psi^m e^{-\phi} \xi\nonumber  \\
 & +\;
   C_L^{\alpha}C_L^{\beta}\Gamma^m_{\alpha\beta}{\partial\over\partial C^m_L} +
   C_L^m\oint \partial X^m +
   C_L^{\alpha}\oint
   \left(
         p_{\alpha}
         - {1\over 2} \partial X^m\Gamma^m_{\alpha\beta}\theta^{\beta}
         - {1\over 24}
         \Gamma^m_{\alpha\beta}\theta^{\beta}
         \theta^{\gamma}\Gamma_{\gamma\delta}^m\partial\theta^{\delta}
         \;
         \right)\nonumber \end{align}
The dependence of the $C$-ghosts corresponds to the strict action.
The coefficients of $C^a$ are the supersymmetry currents  \cite{Siegel:1985xj},
including the translation current $\partial X^m$.
(We must stress, once again, that $C^a$ are constants, they are not worldsheet fields.)

This is {\bf almost} the B-RNS-GSS model. The actual B-RNS-GSS model also consists a quartet
of auxliary ``non-minimal'' fields with the action:
\begin{equation}\int d^2z \left(
                \overline{w}^{\alpha}\overline{\partial} \overline{\lambda}_{\alpha}
                +
                s^{\alpha}\overline{\partial}r_{\alpha}
                \right)\end{equation}
and the contribution to the BRST operator:
\begin{equation}\oint r_{\alpha}\overline{w}^{\alpha}\end{equation}
This additional BRST quartet does not play an important role in the similarity transformation
which we have studied here. In $e^{\iota_{\bf a}}$, the contraction $\iota_{\bf a}$ has an additional term:
\begin{equation}\iota_{\bf a} = \oint c (
                         \partial \theta^{\alpha} \Omega_{\alpha}
                         + \partial \overline{\lambda}_{\alpha} s^{\alpha}
                         )\end{equation}
This additional quartet plays an important role in the second step of the B-RNS-GSS procedure,
which leads to the pure spinor formalism. That step, too, is done by several similarity
transformations, after which some other quartets decouple. Integrating them out leads to the pure spinor model.
This, however, is beyond the scope of this paper.

Our similarity transformation $Re^{\iota_{\bf a}}$ is a refinement of the one used
in \cite{Berkovits:2021xwh}, because we have introduced background ghosts $C_L$.
The similarity transformation of \cite{Berkovits:2021xwh} can be obtained (at the first order in $\theta$)
by putting $C_L$ to zero.
The similarity transformation may be useful for the computation of the string amplitudes, especially since
the naive prescription does not work \cite{Azevedo:2025con}. Extension of the BRST operator
by background ghosts (Eq. (\ref{IntroLinfty})) would help to keep track of the symmetries.

\section{Acknowledgments}\label{Acknowledments}

We would like to thank Nathan Berkovits, Michael Movshev and Nikita Nekrasov for discussions.
We are greatful to SIGMA referees for suggesting references on $L_{\infty}$-actions,
and especially for reminding us about \cite{Elliott:2022brv}, and other useful comments.
This work was supported in part by  FAPESP  grant 2019/21281-4,
and in part by FAPESP grant 2021/14335-0,
and in part by CNPq grant ``Produtividade em Pesquisa'' 307191/2022-2.

\providecommand{\href}[2]{#2}\begingroup\raggedright\endgroup

\end{document}